\documentclass[twocolumn,aps,floats,superscriptaddress]{revtex4}
\usepackage{amsmath}
\usepackage{graphicx,epsfig}

\newcommand{\iomn}{i \omega_n}

\newcommand{\mr}[1]{{{\mathrm{#1}}}}
\newcommand{\ek}{\epsilon_{k}}
\newcommand{\ckc}{c_{k,\sigma}^{\dagger}}
\newcommand{\ck}{c_{k,\sigma}}

\newcommand{\ckp}{c_{k',\sigma'}}

\newcommand{\dsc}{d_{\sigma}^{\dagger}}
\newcommand{\ds}{d_{\sigma}}
\newcommand{\fsc}{f_{\sigma}^{\dagger}}
\newcommand{\fs}{f_{\sigma}}
\newcommand{\fspc}{f_{\sigma '}^{\dagger}}
\newcommand{\fsp}{f_{\sigma '}}

\newcommand{\s}{\sigma}
\newcommand{\w}{\omega}

\newcommand{\kL}{k \in \mr{L}}
\newcommand{\kH}{k \in \mr{H^\pm}}

\newcommand{\VL}{V_{\mr{L}}}
\newcommand{\VH}{V_{\mr{H}}}
\newcommand{\mut}{\widetilde{\mu}}

\def\GroupeEquations#1{\begin{subequations}  #1  \end{subequations}}
\def\moy#1{\left\langle #1 \right\rangle}

\def\parent#1{\left(#1\right)}

\def\ket#1{\left\vert #1 \right\rangle}
\def\bra#1{\left\langle #1 \right\vert}
\def\parent#1{\left(#1\right)}

\renewcommand{\atop}[2]{%
\genfrac{}{}{0pt}{}{#1}{#2}}

\def\ve{\varepsilon}

\def\iomn{i\omega_n}
\def\TC{T_{\text{MIT}}}
\def\UC{U_{\text{MIT}}}

\begin {document}
\bibliographystyle {plain}

\vspace{0.5cm}

\title{Mott transition at large orbital degeneracy:
dynamical mean-field theory}

\author{S. Florens}
\affiliation{Laboratoire de Physique Th{\'e}orique, Ecole Normale
Sup{\'e}rieure, 24 rue Lhomond, 75231 Paris Cedex 05, France \\ and
Laboratoire de Physique des Solides, Universit{\'e} Paris-Sud,
B{\^a}t.~510, 91405 Orsay, France}

\author{A. Georges}

\affiliation{Laboratoire de Physique Th{\'e}orique, Ecole Normale
Sup{\'e}rieure, 24 rue Lhomond, 75231 Paris Cedex 05, France \\ and
Laboratoire de Physique des Solides, Universit{\'e} Paris-Sud,
B{\^a}t.~510, 91405 Orsay, France}

\author{G. Kotliar}
\affiliation{Center for Materials Theory, Department of Physics
and Astronomy, Rutgers University, Piscataway, NJ 08854, USA}

\author{O. Parcollet}
\affiliation{Service de Physique Th{\'e}orique, CEA Saclay,
91191 Gif-Sur-Yvette, France}

\date{May 13th, 2002}

\begin{abstract}
\vspace{0.5cm}
 We study analytically the Mott transition of the $N$-orbital Hubbard
model using dynamical mean-field theory and a low-energy
projection onto an effective Kondo model. It is demonstrated that
the critical interaction at which the insulator appears ($U_{c1}$)
and the one at which the metal becomes unstable ($U_{c2}$) have
different dependence on the number of orbitals as the latter
becomes large: $U_{c1}\propto \sqrt{N}$ while $U_{c2}\propto
N$. An exact analytical determination of the critical
coupling $U_{c2}/N$ is obtained in the large-$N$ limit. The metallic
solution close to this critical coupling has many similarities at
low-energy with the results of slave boson approximations, to
which a comparison is made. We also discuss how the critical temperature
associated with the Mott critical endpoint depends on
the number of orbitals.

\end{abstract}

\maketitle

\section{Introduction}

The Mott transition is one of the central problems in the field of
strongly correlated electronic systems. It is directly relevant to many
compounds, such as $\mbox{V}_2\mbox{O}_3$, $\mbox{NiS}_{2-x}\mbox{Se}_x$
\cite{imada-review}, fullerenes \cite{gunnarsson-fullerenes}, and two-dimensional
organics of the $\kappa$-BEDT family \cite{lefebvre}.
Viewed from a more general perspective,
the Mott transition confronts us with the need to describe electrons
in a solid in an intermediate coupling regime, where perturbative methods
around either the localised or the itinerant limit are of very limited use.
The Mott phenomenon is also a key issue for the development of electronic structure
calculations of strongly correlated materials, a subject of current intense
investigation \cite{held-review}.

A breakthrough in the understanding of this problem resulted from the
development of dynamical mean-field theory (DMFT) \cite{large-d,IPT,RMP,pruschke-review}.
Following the observation that a Mott transition as a function
of the interaction strength $U$ takes place in a (fully frustrated)
one-band Hubbard model at half-filling \cite{ref-QMC} within DMFT,
a rather surprising scenario was unravelled \cite{Mott_scenario}.
In the parameter space defined by temperature and interaction strength, two
kinds of solutions of the DMFT equations were found, corresponding to an
insulator and a metal, with a region $U_{c1}(T) <U<U_{c2}(T)$ in which both
types of solution coexist. The spectral function of the insulating solution has
only high-energy features corresponding to Hubbard bands, while the metallic solution
also has a low-energy quasiparticle resonance. Coexistence results in a
{\it first-order} phase transition at finite temperature.
The first order line ends into a second-order critical
endpoint at $T=\TC$, $U=\UC$ at which the coexistence domain closes
($U_{c1}(\TC)=U_{c2}(\TC)=\UC$). At $T=0$, the transition is also second-order,
because the metallic solution always has lower energy than the insulating one.
A qualitative picture of the evolution of the spectral function
as a function of control parameters was obtained, and this resulted in
successful experimental predictions \cite{optics}.

Early work on these issues relied heavily on an accurate but approximate
technique, the iterative perturbation theory \cite{IPT,Mott_scenario}.
Later on, exact results
near $U_{c2}(T=0)$ were obtained \cite{Moeller,MoellerPHD}, thanks to the development
of a renormalization method exploiting the separation of energy scales
that holds true close to this critical coupling. This method performs a
projection onto a low-energy effective theory after elimination of the
degrees of freedom corresponding to Hubbard bands. A clear picture of
the structure of the DMFT equations near the Mott transition and of the
different critical points was finally obtained through the introduction of a
Landau functional approach \cite{Landau-functionnal}.
These established results for the one-band Hubbard model together with numerical 
methods for solving DMFT equations, such as Quantum Monte Carlo \cite{ref-QMC,QMC-oudo}
and Numerical Renormalization Group \cite{Bulla-NRG}, provide now a consistent
picture of the Mott transition within DMFT.

In this paper, we investigate the Mott transition in the limit of large
orbital degeneracy, within DMFT.
We establish analytically that the critical couplings $U_{c1}(T=0)$
and $U_{c2}(T=0)$ are not only different, but that they have very different scalings
as a function of the number $N$ of orbitals. Indeed, $U_{c1}$ increases as $\sqrt{N}$,
while $U_{c2}$ increases as $N$, so that the coexistence region widens.
Close to $U_{c2}$, the separation of scales which is at the heart of the
projective method becomes asymptotically exact for large $N$. This allows us to
determine exactly the value of the critical coupling $U_{c2}/N$ at large $N$,
and also to demonstrate that slave-boson like approximations become asymptotically valid
at low energy in this limit. Finally, we discuss how the finite temperature metal-insulator
transition depends on the number of orbitals.

Our findings on the critical couplings are in agreement with early QMC results for the
two-orbital model ($N=4$) \cite{multiband-QMC-marcelo}, and with recent results for
higher values of $N$ \cite{Amadon-QMC}. A widening of the coexistence window
$[U_{c1},U_{c2}]$ as $N$ increases is clearly seen in these simulations. Also,
for fixed $N$, the critical coupling required to enter the Mott state is largest \cite{dos}
at the particle-hole symmetric filling $n=1/2$.
The present work also puts the results of Refs.\cite{multiband-QMC-gun,multiband-QMC-han}
in a new perspective: there, a $\sqrt{N}$ scaling of the critical coupling was proposed.
We find that this indeed applies to the coupling where the insulator becomes unstable
($U_{c1}$), while the true $T=0$ Mott transition (at which the quasiparticle residue vanishes)
takes place at $U_{c2}\propto N$. Indeed, in Ref.\cite{multiband-QMC-gun}, the $\sqrt{N}$ scaling
was rationalized on the basis of a stability argument for the insulator. Finally, the rather high
temperatures considered in \cite{multiband-QMC-han} explain
why only the $\sqrt{N}$ dependence was reported there: distinguishing $U_{c2}$
from $U_{c1}$ requires significantly lower temperatures \cite{Amadon-QMC}.

The limit of large orbital degeneracy is relevant to the physics of
systems with partially filled d- or f- shells, as well as to fullerenes.
Direct numerical approaches become prohibitively difficult as
the number of orbital increases (QMC methods scale as a power law of $N$,
while exact diagonalizations \cite{exact-diag} scale exponentially).
It is therefore important for future research to develop approximate but accurate 
impurity solvers which can handle many orbitals. The controlled results that 
we establish in this paper can be used as tests of these approximation methods.

\section{Multi-orbital Hubbard model}
\label{model}

We consider a generalized Hubbard model involving $N$ species of
electrons, with Hamiltonian:
\begin{eqnarray}
\nonumber
H & = & -\sum_{i,j}\sum_{\sigma=1}^{N}\,t_{ij}
d_{i\sigma}^{\dagger}d_{j\sigma}\,+\,\frac{U}{2} \sum_i\left[
\sum_{\sigma=1}^{N}\, \left(d_{i\sigma}^{\dagger}d_{i\sigma} - n\right)
\right] ^2 \\
& - & {\mut }
\sum_i  \sum_{\sigma=1}^{N}\, d_{i\sigma}^{\dagger}d_{i\sigma}
\label{Hubbard}
\end{eqnarray}
where $i,j$ are sites indices, $\sigma $ the orbital index, $\mut $ is
the chemical potential and $n$ the average density of
particles per species :
\begin{equation}
n=\frac{1}{N}\sum_\sigma\langle\,d_{i\sigma}^{\dagger}d_{i\sigma}\rangle
\end{equation}
Introducing $n$ in the Hamiltonian is just a convention for
the chemical potential $\mut$. In particular, it is convenient at
half-filling where the particle-hole symmetry implies 
 $n=\frac{1}{2}$ and $\mut=0$.
For a single site (atom), the
spectrum consists of $N+1$ levels, with energies
$\frac{U}{2}(Q-nN)^2$ depending only on the total charge on
the orbital: $Q=0,\cdots,N$ and with degeneracies
$\binom{N}{Q}$.

The usual single-orbital Hubbard model corresponds to
$N=2$ ($\sigma=\uparrow,\downarrow$). The Hamiltonian
considered here has a full SU($N$) symmetry which includes both spin
and orbital degrees of freedom. Starting from a more realistic
model which assumes an interaction matrix $U_{mm'}$ between
opposite spins and $U_{mm'}-J_{mm'}$ between parallel spins, it
corresponds to the limit of isotropic $U_{mm'}=U$ and vanishing
Hund's coupling $J_{mm'}=0$.

When $U$ is large enough, we expect a Mott insulating state to
exist at fillings $n=Q/N$, corresponding to an integer occupancy
of each site on average $Q=1,\cdots,N-1$.
In this paper, we investigate analytically the nature of these
Mott transitions within DMFT.
We consider only phases with no magnetic ordering and study the
transition between a paramagnetic metal and a paramagnetic Mott
insulator.
In Section \ref{Sec.NGrand}, we extract the large-$N$ behaviour of
$U_{c1}$ and $U_{c2}$ using the low-energy projective technique
analysis of the DMFT equations introduced in
\cite{Moeller} for $N=2$ and extended in \cite{Kajueter-degeneracy} :
we show that the equations for the $U_{c}$'s derived by this method are
greatly simplified for $N\rightarrow \infty$ in the sense that an
atomic limit becomes exact. In section \ref{sec:slave}, we find
quantitative agreement between these results and a
multiorbital slave boson method. Finally, in Sec.\ref{sec:finiteT}, we
consider the Mott transition at finite temperature.

\section{Large-$N$ behaviour of $U_{c1}$ and $U_{c2}$}\label{Sec.NGrand}

\subsection{DMFT and the low-energy projective method}

Let us recall briefly the DMFT equations and their low-energy
projective analysis \cite{Moeller,MoellerPHD}. DMFT maps the lattice
Hamiltonian above onto a multi-orbital Anderson impurity model
with the same local interaction term than in (\ref{Hubbard}) and a
hybridization function $\Delta(\iomn)$. The local Green's function
reads: $G_d(\iomn)^{-1}=\iomn+\mut-\Delta(\iomn)-\Sigma_d(\iomn)$.
In this expression, $\mut$ is the chemical potential (shifted by
the Hartree contribution) and $\Sigma_d(\iomn)$ is a local
self-energy. A self-consistency requirement is imposed, which
identifies $G_d(\iomn)$ with the on-site Green's function of the
lattice model with the same self-energy $\Sigma_d(\iomn)$. This reads:
\begin{equation}
G_d(\iomn) = \int \mr{d} \ve\, \frac{ D(\ve)}
{ i\omega_{n} + \mut - \epsilon - \Sigma_{d} (i\omega_{n})}
%= \int \mr{d} \ve\, \frac{ D(\ve)} { \Delta(\iomn) +
%G_d^{-1} (\iomn) - \epsilon }
 \label{eq:self-consistency}
\end{equation}
In this expression, $D(\ve)$ is the free electron density of state
(d.o.s.) corresponding to the Fourier transform of the hopping
matrix elements $t_{ij}$. In the particular case of a
semi-circular d.o.s. with half-width $D=2t$, 
equation~(\ref{eq:self-consistency}) simplifies to:
\begin{equation}
\Delta(\iomn)\,=\, t^{2}\,G_d(\iomn)
 \label{eq:self-consistency:scdos}
\end{equation}
Solving the impurity model subject to (\ref{eq:self-consistency})
determines both the hybridization function and the local Green's
function in a self-consistent manner. In order to give an explicit
Hamiltonian form to the Anderson impurity model, the function $\Delta$
can be represented with an auxiliary bath of conduction
electrons $c_{k\sigma}$, with effective single-particle energies $\ek$
and effective hybridizations $V_k$ to the local orbital such that  :
\begin{gather}\label{eq:anderson}
\Delta(i \w) \equiv \sum_{k} \frac{|V_k|^2}{i \w - \ek}
\\
\nonumber
H =\sum_{k\sigma }\ek \ckc \ck +
\sum_{k,\s} V_{k} ( \ckc \ds + \dsc \ck ) \\
-\mut \,\sum_\sigma \dsc\ds\, + \frac{U}{2}\left[ \sum_{\s} (\dsc \ds - n) \right] ^2
\end{gather}

Within DMFT, the Mott transition at $U_{c2}$ and $T=0$ is associated with a
separation of energy scales. For $U$ slightly smaller than
$U_{c2}$, the d-electron spectral function has a sharp
quasiparticle peak, well separated from the high-energy Hubbard
bands. Naturally, a small amount of spectral weight does connect
these two features in the metal. Slightly above $U_{c2}$, the Mott
gap $\Delta_g$ separating the Hubbard bands of the insulating
solution is finite. While the weight $Z$ of the quasiparticle peak
vanishes at $U_{c2}$, the Mott gap $\Delta_g$ disappears only at the
smaller critical coupling $U_{c1}$. Because of the
self-consistency condition (\ref{eq:self-consistency}), the
hybridization function of the effective conduction bath (and hence
the $\ek$'s) also displays this separation of scales.

This remark led the authors of Ref. \cite{Moeller} to use a {\it
projective method} in order to study the critical behaviour at
$U_{c2}$. Within this approach, the effective conduction electron
bath is separated into a high energy sector ($k\in\,\mr{H}^\pm$)
associated with the upper ($\mr{H}^+$) and lower ($\mr{H}^-$) Hubbard bands,
and a low-energy sector ($k\in\,\mr{L}$) associated with the
quasiparticle resonance. The impurity model Hamiltonian is thus
written as (for simplicity we take $\VL$ and $\VH$ independent of $k$):
\GroupeEquations{
\begin{align}
H  &= H_{\mr{L}} + H_{\mr{H}} + H_{\mr{HL}}\\
H_{\mr{L}} &= \sum_{\kL, \s} \ek \ckc \ck
\\
\nonumber
H_{\mr{H}} &=  \sum_{\kH, \s} \left[ \ek \ckc \ck
    + \VH  \ckc \ds + h.c. \right] \\
    & + \frac{U}{2} \left[ \sum_{\s} \dsc \ds - n \right] ^2 -
 \mut  \sum_{\s} \dsc \ds
\\
H_{\mr{HL}} &= \sum_{\kL, \s} \left[   \VL \ckc \ds + h.c.  \right]
\end{align} }
We note that the self-consistency condition (\ref{eq:self-consistency})
implies that $\sum_{k\in \mr{L}}\VL^2/t^2$ is proportional to the
quasi-particle residue $Z$, so that it is clear that the low-energy hybridization
vanishes at $U_{c2}$.

In order to obtain an effective theory close
to $U_{c2}$, one first diagonalizes the high-energy part $H_{\mr{H}}$ of
the Hamiltonian, and obtains an effective low-energy hamiltonian
by expanding in $\VL$.
From the presence of a finite gap in the high-energy bath
$\epsilon_k \, (\mr{for} \; k\in\mr{H^{\pm}})$, $H_\mr{H}$ describes
an impurity in a semiconducting bath. The position of the d-level ({\it i.e.} the
chemical potential) fixes the number of electrons in the
ground-state. For a charge $Q=nN$, this high-energy problem has
a $\binom{N}{Q}$ degenerate ground-state, spanned by
eigenvectors $|\mu\rangle$ (with eigen-energy $\mr{E_{gs}}$).
The detailed form of these states is
complicated (they mix with higher impurity charge states through
$\VH$), but {\it they form a representation of the SU($N$) ``spin''
group, of dimension $\binom{N}{Q}$}.

Using a generalized Schrieffer-Wolff (SW) transformation to order
$\VL^2$ in order to project onto the low-energy Hilbert space (defined by the
selected charge sector of the impurity $d$,
and the low-energy sector for the conduction bath $\kL$), one obtains an effective Kondo
Hamiltonian \cite{Moeller,Kajueter-degeneracy,MoellerPHD} :
\begin{equation}
\label{eq:kondo-general}
H_{\mr{eff}} = \sum_{k\in\,\mr{L},\sigma} \ek
c^\dagger_{k,\sigma}c_{k,\sigma} \,+\,
\frac{\VL^{2}}{U}\sum_{kk',\mu\nu,\sigma\sigma'}\, J_{\mu\nu}^{\sigma\sigma'}
X^{\mu\nu}\,c^\dagger_{k,\sigma}c_{k',\sigma'}
\end{equation}
where $X^{\mu\nu}\equiv |\mu\rangle\langle\nu|$ is a
Hubbard operator connecting two states in the ground state manifold.
The Hamiltonian contains potential scattering terms for
$\sigma=\sigma'$ and spin-flip (Kondo) terms for
$\sigma\neq\sigma'$, with the spin operator
$S_{\mu \nu } = X^{\mu\nu}  - \delta _{\mu \nu }X^{\rho \rho }$.
 The matrix of (dimensionless) coupling constants reads
\cite{Kajueter-degeneracy,MoellerPHD}:
\begin{equation}\label{DefJ}
J^{\s \s'}_{ \mu  \nu } \equiv
\moy{\mu \left| d^{\dagger}_{\s' }
\frac{U}{H_{\mr{H}}-\mr{E}_{gs}} d_{\s }  \right| \nu  }
-
\moy{\mu \left| d_{\s }
\frac{U}{H_{\mr{H}}-\mr{E}_{gs}} d^{\dagger}_{\s'}  \right| \nu
}
\end{equation}
We emphasize that the low-energy Hamiltonian has been derived to first order in
$\VL^2$: this is asymptotically justified exactly at the critical
point (and hence sufficient to determine the critical coupling),
but deviations away from the critical point require to consider
higher order terms in $\VL$.
Finally, we note that the bandwidth of this Kondo problem is of order
$Z\,t$, while the coupling is of order $\VL^2/U\sim Z\,t^2/U$.
Hence, the dimensionless coupling constant of the effective
low-energy Kondo problem is of order one: we have an
{\it intermediate coupling} Kondo problem \cite{Landau-functionnal}.

The conditions that determine $U_{c1}$ and $U_{c2}$  have been derived
within the projective method \cite{Moeller,Kajueter-degeneracy} and
within the more general Landau functional approach \cite{Landau-functionnal}
(in which the stability of the insulating solution to different kinds of
low-energy perturbations is studied using a Landau functional of the
hybridization function $\Delta(\iomn)$). The condition for $U_{c2}$ reads:
\begin{eqnarray}\label{UCondition}
\frac{U_{c2}^{2}}{t^{2}} & = &  \sum_{ \alpha \beta \mu \nu}
%{1 \leq \alpha \beta \leq N\atop 1\leq  \mu \nu\leq d (n,N) }
 \binom{N}{nN}^{-1} \,
|J^{\alpha \beta }_{\mu \nu }|^{2} \\
\nonumber
& + & \moy{X_{\mu \nu } c^{\dagger }_{\alpha } c_{\beta } }
\biggl (
J^{\gamma \beta }_{\rho \nu }{J^{\gamma \alpha }_{\rho \mu}}^{*} -
J^{\gamma \beta }_{\mu \rho  } {J^{\gamma \alpha }_{\nu \rho}}^{*}
\biggr)
\end{eqnarray}
As explained in \cite{Landau-functionnal,Kajueter-degeneracy}, an
analogous equation is found for $U_{c1}$, where off-diagonal terms in
the average are  taken to zero.
For finite $N$, this equation is complicated to solve, since
the average on the right-hand side is a function of $J$ which  we can only
obtain by solving the low energy Kondo problem explicitly.
However, we will show in the following that :
\begin{enumerate}
\item Using the $SU (N)$ symmetry, the tensor structure of $J$ can be
parametrized with only two numbers $A$
and $B$, which are dimensionless functions of $N$, $n$,
and $\VH/U$. This greatly simplifies
(\ref{UCondition}) (Section \ref{sub.simpliftensor}).
\item An explicit calculation of $A$ and $B$ can be performed in the atomic limit
$\VH=0$. This allows us to show that the two
critical couplings have the following $N$-dependence at large $N$:
\GroupeEquations{
\begin{align}\label{ScalingU}
U_{c1} &=  \sqrt{N} \, \widetilde{U}_{c1}+\cdots \\
\label{ScalingU-b}
U_{c2} &= N  \, \widetilde{U}_{c2}+\cdots
\end{align}}
In fact, the atomic limit evaluation of $A$ and $B$ becomes asymptotically
exact when $U$ is taken to be proportional to $N$.
This is not the case when $U\propto\sqrt{N}$,
but we show that corrections to the atomic limit
do not modify the scaling (\ref{ScalingU}).

\item For $U\propto N$, the effective Kondo problem can actually be solved
explicitly at large $N$, using standard techniques. This allows for an explicit
computation of the matrix element $\moy{X_{\mu \sigma }
c^{\dagger }_{\alpha } c_{\beta } }$ involved in (\ref{UCondition}).
Since, from the above remark, the effective couplings entering the Kondo
problem can be evaluated in the atomic limit in that case, this allows us
to determine the prefactor $\widetilde{U}_{c2}$ exactly
(Sec.~\ref{sub:largeNKondoeff}). Because the atomic limit is not asymptotically exact
for $U\propto\sqrt{N}$, we are not able to obtain an exact determination of
$\widetilde{U}_{c1}$.

\end{enumerate}

\subsection{Simplication due to the $SU (N)$ symmetry}\label{sub.simpliftensor}

Computing the couplings (\ref{DefJ}) is a complicated task in general,
since it requires a knowledge of the states $|\mu \rangle , |\nu \rangle $, and thus
a diagonalization of $H_{\mr{H}}$. However, general considerations
based on the SU($N$) symmetry of the model
allow one to reduce the unknown Kondo couplings $J^{\s \s'}_{ \mu \sigma }$ to two dimensionless
numbers only.
Let us consider the operator:
\begin{align}\label{DefO}
O_{1}^{\s \s' } \equiv d_{\s}
\frac{1}{H_{\mr{H}}-\mr{E}_{gs}} d^{\dagger}_{\s' }
\end{align}
Since the Hamiltonian is SU($N$) symmetric, this operator
transforms under SU($N$) with the representation $F \otimes F^{*} = id
\oplus ad $ :
the tensor product of the fundamental $F$ (one box in the Young
tableau language) and its complex conjugate $F^{*}$ ($N-1$ boxes), which reduces to the sum of the
identity $id$ and the adjoint $ad$.
If we denote by $R$ the ground state representation (a column
with $Q=nN$ boxes), the tensor structure of $J$ is then given by the
Wigner-Eckart theorem : we have to find the occurence of $R$ in
 $(id \oplus ad)\otimes R$.  We have a trivial term $id\otimes R=R$ and
{\it a single} non trivial one since $ad\otimes R= R\oplus \dots $,
where $\dots $ denotes other irreducible representations.
Then we have :
\begin{equation}
\moy{\nu \left| d_{\s}
\frac{1}{H_{\mr{H}}-\mr{E}_{gs}} d^{\dagger}_{\s' }
\right|\mu  } = A_{1}
\delta_{\s \s'}\delta_{\mu \nu  } + B_{1}C^{\s
\s'}_{\mu \nu }
\end{equation}
where $C$ are the Clebsch-Gordan coefficients $\moy{R | ad\otimes
R}$.

In order to avoid the computation of $C$, we can use the following
argument.
Let us consider the operator  $d_{\s }  d^{\dagger}_{\s' }$, which
has the same symmetry as $O_{1}$, so the previous symmetry argument
gives
\begin{equation}
\moy{\nu \left| d_{\s}  d^{\dagger}_{\s' }\right|\mu } = A_{2}
\delta_{\s \s'}\delta_{\mu \nu  } + B_{2}C^{\s \s'}_{\mu \nu }
\end{equation}
with the {\it same} $C$, but {\it different  coefficients} $A_{2},B_{2}$. Since $B_{2}\neq
0$,  we can eliminate $C$ between the two equations. We finally obtain the general form,
valid to first order in $\VL^2$ for a $SU(N)$ symmetric model:
\begin{equation}\label{Jreduced}
J^{\s \s' }_{ \mu  \nu } =
A \,\delta_{\s\s' }\delta_{\mu \nu  } + B\, \moy{\mu\left| d^{\dagger}_{\s' }
d_{\s }\right|\nu  }
\end{equation}
In this expression, $A\left(N,n;\VH/U\right)$ and
$B\left(N,n;\VH/U\right)$ are dimensionless coefficients
which depend only on the dimensionless ratio $\VH/U$ associated
with the high-energy problem (and of course a priori also on
orbital number $N$ and density $n$).
%The last bracket can  be computed straightforwardly (using free
%fermions), but we do not need to do this computation explicitly :

Inserting (\ref{Jreduced}) into equation (\ref{UCondition}),
we obtain the simplified form of the conditions which determine the critical
couplings:
\GroupeEquations{
\begin{align}\label{ReducedUCondition}
\frac{U_{c1}^2}{t^{2}} &= \parent{A + n\,B}^{2} + B^{2}\,n(1-n)(N+1)
\\
\label{ReducedUCondition-b}
\nonumber
\frac{U_{c2}^2}{t^{2}} &= \parent{A + n\,B}^{2} + B^{2}
\,n(1-n)(N+1) \\
& - B^{2} \sum_{\s \s'}\moy{S_{\s \s'} {c}_{\s'}^{\dagger} {c}_{\s}}
\end{align}
}

\subsection{The atomic limit}\label{sub:limitatomicexact}

The atomic limit approximation is defined by $\VH=0$. In this limit,
the states $|\mu \rangle $ are just the impurity states $|\mu
\rangle_{0}\equiv |\sigma_1\cdots\sigma_Q\rangle$ ($Q=0,\cdots,N$),
without mixing with the high-energy electron  sector.
In that case, $A$ and $B$ (from (\ref{Jreduced})) can be computed
explicitly (~we choose here $\mut=0$, {\it i.e.} the chemical potential is
bound to be at the center of a step in the Coulomb staircase). We find:
\begin{equation}\label{ValueABAtomiclimit}
A=-2, B= 4
\end{equation}
We note that these coefficients do not depend on $N$ in the atomic limit.
Moreover, it is clear that, in general (for an explicit proof,
see Sec.\ref{sub:largeNKondoeff}) :
\begin{equation}\label{aprioriscalingbraket}
\sum_{\s \s'}\moy{S_{\s \s'}
{c}_{\s'}^{\dagger} {c}_{\s}} \propto N^{2}
\end{equation}
When using these results into
(\ref{ReducedUCondition}-\ref{ReducedUCondition-b}), we obtain the
$N$-dependence of the critical couplings quoted in (\ref{ScalingU}-\ref{ScalingU-b}).

This derivation relies on the use of the atomic limit $\VH=0$.
An important question is therefore to check that the $N$-dependence
of the critical couplings is not modified by
the interaction $\VH$ that dresses the atomic states $|\mu\rangle$.
We examine carefully this question in Appendix~\ref{appendix}, where we
give an explicit proof that this is indeed the case.
The problem is in fact quite different for $U_{c1}$ and $U_{c2}$.
For the latter, $U$ must be scaled as $U=N\widetilde{U}$. In that case, we
show in the appendix that the atomic limit calculation of $A$ and $B$ is in
fact {\it exact} at large $N$. This can be
expected since the band gap diverges with $N$, so that the Hubbard bands
are well separated from the low energy degrees of freedom. However, one might
worry that the growing number of orbitals ($N$) could counterbalance this
effect. This is actually not the case, and we show that
the corrections to $J$ (or $A$,$B$) coming from $\VH$
are subdominant in large-$N$, order by order in perturbation theory.
Moreover, the low-energy effective Kondo problem becomes exactly solvable
when $U\propto N$ (i.e close to $U_{c2}$): this is the subject of the next
section (Sec.~\ref{sub:largeNKondoeff}), in which the exact value of the
prefactor $\widetilde{U}_{c2}$ is obtained.

The problem is more involved for $U_{c1}$, corresponding to the scaling $U\propto\sqrt{N}$.
In that case, corrections in $\VH$ beyond the atomic limit do produce corrections to
$A$ and $B$ (hence to $J$). However, we are able to show (Sec.~\ref{sub:Uc1} of the appendix)
that these corrections are of order
one, so that the scaling $U_{c1}=\sqrt{N}\widetilde{U}_{c1}$ does hold.

\subsection{Explicit large-$N$ solution of the effective low energy
Kondo problem at $U_{c2}$}\label{sub:largeNKondoeff}

In this section, we explicitly solve the low energy effective Kondo
problem in the large-$N$ limit using standard methods
\cite{Read-Newns}, in order to obtain the prefactor
$\widetilde{U}_{c2}$.
The effective Hamiltonian (\ref{eq:kondo-general}) reduces at
order ($1/N^2$) to:
\begin{eqnarray}
H_{\mr{eff}} & = & \sum_{\kL,\s} \ek \ckc \ck \\
\nonumber
& & +\; \frac{4 \VL^{2}}{N\widetilde{U}}
\sum_{k k'\in \mr{L},\s \s'}
\left( \fspc\fs - \frac{\delta_{\sigma \sigma '}}{2}
 \right) \ckc \ckp
\label{eq:Heff}
\end{eqnarray}
This model becomes exactly solvable at large $N$ because the Kondo coupling
scales as $1/N$ (which is a result of $U\propto N$).
Introducing an auxiliary boson field $b$ conjugate to
$\sum_{k\sigma}f^\dagger_\sigma\,c_{k\sigma}$, and a Lagrange
multiplier field $\lambda$ to enforce the constraint $\sum_{\s}
\fsc\fs=Nn$, we see that, as $N\rightarrow\infty$, the field $b$ undergoes
a Bose condensation, and the saddle-point values of the
$b,\lambda$ fields are determined by the equations (at $T=0$):
\GroupeEquations{
\begin{align}
G_f^{-1}
(\omega) &= \omega + \lambda - b^2 G_{c\mr{L}}^0(\omega) \label{eq:col}
\\
n &= -\frac{1}{\pi}\int_{-\infty}^0 \mr{d}\w \mbox{Im} G_f(\w) \label{eq:nb-fermion}
\\
\frac{\widetilde{U}_{c2}}{4\VL^2} &=
\int \mr{d}\tau G_f(\tau) G_{c \mr{L}}^0(\tau) =
\frac{1}{\pi} \int_{-\infty}^0 \mr{d}\w \mbox{Im} [G_f(\w) G_{c
\mr{L}}^0(\w)] \label{eq:Uc}
\end{align}
}
In this expression, $G_{c \mr{L}}^0$ denotes the Green's function
of the low-energy part of the effective bath of conduction
electrons, in the absence of Kondo coupling, namely:
\begin{equation}
G_{c \mr{L}}^0(i\w) =\sum_{\kL} \frac{1}{i \w - \ek} =
\frac{1}{\VL^2}\,  \Delta_{\mr{L}} (i \w) \label{eq:bainL}
\end{equation}

First we express the self-consistency condition
(\ref{eq:self-consistency}) or more precisely its low-energy
counterpart. We relate the low-energy part of the on-site Green's function,
$G_{d\mr{L}}(\iomn)$, to the auxiliary-fermion one, $G_f(\iomn)$, by
imposing that the low-energy part of the interacting Green's
function of the effective bath of conduction electrons, $G_{c\mr{L}}$,
can be calculated either from the original Anderson model or from
the low-energy projected Kondo model, with identical results.
$G_{c\mr{L}}$ is related to $G_{c\mr{L}}^0$ through the conduction
electron ${\cal T}$-matrix:
\begin{equation}
G_{c\mr{L}}\,= \,G_{c\mr{L}}^0 + (G_{c\mr{L}}^0)^2\,{\cal T}
\end{equation}
The Anderson model ${\cal T}$-matrix is $\VL^2G_{d\mr{L}}$, while it is
$b^2\,G_f$ for the low-energy Kondo model. Therefore $G_{d\mr{L}}$ is
directly related to $G_f$ by:
\begin{equation}
\VL^2 G_{d\mr{L}}\, =\, b^2 G_f\,\equiv {\cal T}
\label{eq:Tmatrix}
\end{equation}
Using this into
Eq.(\ref{eq:col}), one finds the expression of $G_{d\mr{L}}$ in terms
of the low-energy hybridization $\Delta_\mr{L}$:
\begin{equation}
G_{d\mr{L}}^{-1}(\iomn)=\frac{1}{Z}\,
\left(\iomn+\lambda\right)-\Delta_\mr{L}(\iomn)
\label{imp-solver}
\end{equation}
where we have introduced the notation:
\begin{equation}
Z\,\equiv\,\frac{b^2}{\VL^2}
\end{equation}
We can finally obtain $G_{d\mr{L}}$ in closed form by making use of
(\ref{imp-solver}) into the self-consistency condition
(\ref{eq:self-consistency}) (restricted to the low-energy sector).
This yields:
\begin{eqnarray}
G_{d\mr{L}}(i\w) & = & \int \mr{d} \ve\, \frac{D(\ve)}{
\Delta_\mr{L} (i \w) + G_{d\mr{L}}^{-1} (i \w) - \ve}\\
& = & \int \mr{d} \ve\,\frac{D(\ve)}{\frac{1}{Z}(i\w + \lambda) - \ve}
\label{eq:Lself-consistency}
\end{eqnarray}
The low-energy part of the local spectral density reads:
\begin{equation}
\rho_{d\mr{L}}(\omega) \equiv -\frac{1}{\pi} \mbox{Im} G_{d\mr{L}} (\w +
i0^+) = D \left( \frac{\w + \lambda}{Z} \right)
\label{eq:Ldos}
\end{equation}
By comparing this form of $G_{d\mr{L}}$ to its expression in terms of the
local self-energy, we recognize that $Z$ is the quasi-particle
weight and $\lambda$ yields the zero-frequency limit of
$\Sigma_d$:
\begin{equation}\label{eq:Fermi-liquid}
\mu-\Sigma_d(0)=\frac{\lambda}{Z}\,\,\,,\,\,\,
Z\,=\,\left[1-\frac{\partial\Sigma_d}
{\partial\omega}|_{\omega=0}\right]^{-1}
\end{equation}
The value of $\lambda/Z$ can be determined by using the explicit
form of $\rho_d=Z\rho_f$ into (\ref{eq:nb-fermion}):
\begin{equation}
n = \int_{-\infty}^{0} d\w\,\frac{1}{Z}\,D \left( \frac{\w +
\lambda}{Z} \right)
\end{equation}
Changing variables, this leads to $\lambda /Z = \mu_0(n)$
where $\mu_0(n)$ is the non-interacting value of the chemical
potential, defined by:
\begin{equation}
n\,=\,\int_{-\infty}^{\mu_0} d\ve\,D(\ve)
\end{equation}
We note that this is precisely the result expected from
Luttinger's theorem (we have  $\mu-\Sigma_d(0)=\mu_0(n)$
which  insures that the Fermi surface volume
is unchanged by interactions). Hence, the low-energy part of the
spectral function (quasi-particle resonance) is simply given by:
\begin{equation}
\rho_{d\mr{L}}(\omega) = D \left( \frac{\w}{Z} +\mu_0(n) \right)
\label{eq:Ldosfinal}
\end{equation}
Its shape is simply given by the non-interacting d.o.s., centered
at $\omega=-Z\mu_0$ and with a width renormalized by $Z$.

Second, we can now use (\ref{eq:Uc}) and
$\mbox{Im}[\Delta_LG_f]=\frac{\omega+\lambda}{Z}\,\mbox{Im}G_f$,
to obtain :
\begin{eqnarray}
\frac{\widetilde{U}_{c2}}{4} & = & - \frac{1}{Z^2}\,\int_{-\infty}^{0} \mr{d}
\w (\w + \lambda) \rho_{dL}(\w) \\
& = & - \,\int_{-\infty}^{0}
\frac{\mr{d}\w}{Z}\frac{\w + \lambda}{Z}\, D \left( \frac{\w +
\lambda}{Z} \right)
\end{eqnarray}
Clearly, $Z$ {\it drops out of this equation}, which has therefore a
unique solution, the critical coupling $\widetilde{U}_{c2}$ at which $Z$ vanishes.
This is expected: the low-energy effective Hamiltonian is valid
only exactly at the critical point. A determination of the
critical behaviour of $Z$ slightly away from the critical point
would require considering higher-order terms in the effective
low-energy Hamiltonian. We therefore obtain the large-$N$ value of
the critical coupling (using the expression of $\lambda/Z$ found
above):
\begin{equation}
\label{uc2-critical}
\widetilde{U}_{c2} = \frac{U_{c2}}{N} = 4\,\varepsilon_0(n)
\end{equation}
with:
\begin{equation}
\label{def-epsilon0}
\varepsilon_0(n) \equiv - \int_{-\infty}^{\mu_0(n)}d\ve\,\ve\,D(\ve)
\end{equation}
This result is reminiscent of the critical coupling obtained in
the Gutzwiller or slave-boson methods \cite{lu}: a detailed
comparison is made in Sec.\ref{sec:slave}.
Taking for example a rectangular d.o.s. of half-width $D$,
Eq.(\ref{uc2-critical}) yields:
\begin{equation}
\frac{U_{c2}}{N}=4D\,n(1-n)
\end{equation}
$\widetilde{U}_{c2}$ is largest at half-filling ($n=1/2$).
This is because orbital fluctuations are maximum there, so that it is
more difficult to destroy the metallic phase (see however \cite{dos}).
Obviously, for finite $N$,
a Mott transition is found only for rational values of
$n=Q/N$ with integer $Q$ (these values are dense
on the interval $[0,1]$ as $N$ becomes large).

Another way to obtain $U_{c2}$ is to use the criterion of section
\ref{sub:limitatomicexact} together with the present large-$N$ limit
in order to
compute the matrix element $\sum_{\s \s'}\moy{S_{\s \s'}
{c}_{\s'}^{\dagger} {c}_{\s}}$. This can be exactly evaluated at $N=\infty$
using the effective action
for the Kondo model (\ref{eq:Heff}):
\begin{eqnarray}
S_{\mr{eff}} & = & \int_0^\beta \mr{d}\tau \sum_\sigma \fsc \partial_\tau \fs -
\lambda \sum_\sigma(\fsc \fs - n) \\
\nonumber
& + & \int_0^\beta \mr{d}\tau \frac{- N\widetilde{U}}{4 \VL^2} b^2
  + \sum_\sigma b \fsc c_\sigma + b c^\dagger_\sigma \fs \\
\nonumber
& - & \int_0^\beta \mr{d}\tau \int_0^\beta \mr{d}\tau' \sum_\sigma
c^\dagger_\sigma(\tau) G_{c\mr{L}}^0(\tau-\tau')^{-1} c_\sigma(\tau')
\end{eqnarray}
Because this action is gaussian, it is straightforward to get the
following average (here $\sigma\neq\sigma'$):
\begin{eqnarray}
\nonumber
& & \left< \fsc \fsp c_{\sigma'}^\dagger c_\sigma \right> =  \\
\nonumber
& & \left< \fsc \fsp
\int_0^\beta \mr{d}\tau G_{c\mr{L}}^0(\tau) b f_{\sigma'}^\dagger (\tau)
\int_0^\beta \mr{d}\tau' G_{c\mr{L}}^0(\tau') b \fs(\tau')\right> \\
& & =  - b^2 \left[ \int_0^\beta \mr{d}\tau G_{c\mr{L}}^0(\tau) G_f(\tau)
\right]^2
\end{eqnarray}
We find therefore:
\begin{equation}
\sum_{\s \s'} \moy{S_{\s \s' } c_{\s' }^{\dagger}c_{\s }} = -
N^2 b^2 \left[ \int \mr{d} \tau G_f(\tau) G_{c
\mr{L}}^0(\tau) \right]^2 + {\mathcal{O}}(N)
\end{equation}
which indeed scales as $N^{2}$.
Inserting this into (\ref{ReducedUCondition-b}), we have :
\begin{equation}
U_{c2} = 4 t \frac{\VL}{\sqrt{Z} t} \, N \, b
\int \mr{d}\tau G_f(\tau) G_{c \mr{L}}^0(\tau)
\end{equation}
which coincides with Eq.~(\ref{eq:Uc}) since $ \sqrt{Z} = b /\VL$.
Hence, we have checked that both methods (explicit large-$N$ solution and
projective criterion of Sec.\ref{sub:limitatomicexact}) yield the same result for the
critical interaction $U_{c2}$.

\section{Comparison to a slave-boson approach}
\label{sec:slave}

In this section, we show that a multiorbital slave boson
approximation reproduces the previous results close to the critical
coupling $U_{c2}$.  This method is an
extension to many orbitals of the Gutzwiller approximation,
formulated in terms of slave bosons by Kotliar and Ruckenstein
\cite{4bosons} and extended to many orbitals in
\cite{fresard-kotliar}.
Indeed, with the scaling $U=N\widetilde{U}$, the Mott gap is of order $N$ and
quasiparticles dominate the physics over most of the energy range.
Furthermore, the expression (\ref{eq:Ldos}) for the physical
spectral function close to $U_{c2}$ contains precisely the same
ingredients than in slave boson methods, namely: a shift of the
effective level which insures Luttinger's theorem, and a
quasiparticle weight $Z$ (vanishing at the transition). The
self-energy contains no further renormalizations (in particular
the quasi-particle lifetime is infinite). This bare-bone picture
of a strongly correlated Fermi liquid is precisely that emphasized
by slave boson approaches.

Following \cite{fresard-kotliar}, one introduces $2^N$ slave
bosons $\phi_{\sigma_{1}\dots \sigma_{p}}$ associated with a given
spin configuration $\sigma_1\cdots\sigma_p$ on the impurity
orbital. The probability that the orbital is in a given spin
configuration is simply given by $\langle
\phi^\dagger_{\sigma_{1}\dots \sigma_{p}}\phi_{\sigma_{1}\dots
\sigma_{p}}\rangle$. The physical electron operator is represented
as:
\begin{equation}
d^{\dagger }_{\alpha }\,=\, z_{\alpha}\,f^{\dagger}_{\alpha}
\end{equation}
with:
\begin{eqnarray}\label{eq:sb-representation}
z_{\alpha } & \equiv & g_1
\left( \sum_{p} \sum_{\alpha \notin (\sigma_{1},\dots ,\sigma_{p})}
\phi^{\dagger}_{\alpha \sigma_{1}\dots \sigma_{p}}
\phi_{\sigma_{1}\dots \sigma_{p}} \right) g_2 \\
\nonumber
g_1  & \equiv &
\left[1-\sum_{p} \sum_{\alpha \in
(\sigma_{1},\dots ,\sigma_{p})} \phi^{\dagger}_{\sigma_{1}\dots
\sigma_{p}} \phi_{\sigma_{1}\dots \sigma_{p}}\right]^{-1/2} \\
\nonumber
g_2 & \equiv &
\left[1-\sum_{p} \sum_{\alpha
\notin (\sigma_{1},\dots ,\sigma_{p})}
\phi^{\dagger}_{\sigma_{1}\dots \sigma_{p}} \phi_{\sigma_{1}\dots
\sigma_{p}}\right]^{-1/2}
\end{eqnarray}
This representation is supplemented by two constraints:
\GroupeEquations{
\begin{align}\label{eq:sb-const1}
f^{\dagger }_{\alpha } f_{\alpha } &= \sum_{p} \sum_{\alpha \in
(\sigma_{1},\dots ,\sigma_{p})} \phi^{\dagger}_{\sigma_{1}\dots
\sigma_{p}} \phi_{\sigma_{1}\dots \sigma_{p}}
\\ \label{eq:sb-const2}
1&= \sum_{p} \sum_{(\sigma_{1}\dots \sigma_{p})}
\phi^{\dagger}_{\sigma_{1}\dots \sigma_{p}} \phi_{\sigma_{1}\dots
\sigma_{p}}
\end{align}}
At this stage, this is an exact representation of the Hilbert space,
which can be used to rewrite the Hubbard Hamiltonian.
In order to obtain an approximate solution, we shall assume that all auxiliary bosons
undergo a Bose-condensation, i.e become static c-numbers, and that furthermore
the corresponding expectation value depends only on the total charge $p$ and not on the
specific spin configuration $\sigma_1,\cdots,\sigma_p$. We shall thus set, at each
site:
\begin {equation}\label{Assumptions}
\phi_{i\sigma_{1}\dots \sigma_{p}}= \phi_{p}
\end{equation}
We emphasize that this approximation {\it does not} correspond to a controlled
saddle-point, even in the large-$N$ limit, since the number of auxiliary fields grows
rapidly with $N$.

Using this Ansatz, the Hubbard Hamiltonian can be rewritten in the form:
\begin{equation}\label{eq:Hubbard-SB}
H = -\sum_{i,j}\sum_{\sigma=1}^{N}\,t_{ij}
z^2\,f_{i\sigma}^{\dagger}f_{j\sigma}\,+\,\frac{U}{2} \sum_i
\sum_{p=0}^{N}\binom{N}{p}\phi_p^2\,(p-Nn)^2
\end{equation}
with:
\begin{equation}
z\,=\,\frac{1}{\sqrt{n(1-n)}}\,\sum_{p=0}^{N-1}\binom{N-1}{p}\phi_p\phi_{p+1}
\end{equation}
and the constraints:
\begin{eqnarray}
\sum_{p=0}^{N}\binom{N}{p}\phi_p^2 & = & 1\\
\sum_{p=1}^{N}\binom{N-1}{p-1}\phi_p^2 & = & n
\equiv\langle f^\dagger_\sigma f_\sigma\rangle
\end{eqnarray}
The free-energy corresponding to this Hamiltonian must then be minimized with
respect to the $N+1$ variational parameters $\phi_p$'s, subject to the two constraints
above. In the following, we briefly outline this procedure, focusing for simplicity
on the half-filled case ($n=1/2$) at zero-temperature.

In this case, particle-hole symmetry makes the analysis simpler. It implies that
$\phi_p=\phi_{N-p}$, and it can then be shown that the constraint
$\sum_{p=1}^{N}\binom{N-1}{p-1}\phi_p^2=n=1/2$ is automatically satisfied when
the first one ($\sum_{p=0}^{N}\binom{N}{p}\phi_p^2=1$) is. Evaluating the ground-state
energy of (\ref{eq:Hubbard-SB}) yields the following expression to be minimized
at $T=0$ (subject to just one constraint):
\begin{eqnarray}
\nonumber
\frac{1}{N} E_0 & = &
-4\varepsilon_0 \left[\sum_{p=0}^{N-1}\binom{N-1}{p}\phi_p\phi_{p+1} \right]^2 \\
& + & \frac{\widetilde{U}}{2}\,\sum_{p=0}^{N}\binom{N}{p}\phi_p^2\,\left(\frac{N}{2}-p\right)^2
\end{eqnarray}
with $\varepsilon_0$ the (absolute value) of the non-interacting kinetic
energy per orbital (as defined in (\ref{def-epsilon0})), and $\widetilde{U}\equiv U/N$, as defined above.

It is clear from the Hamiltonian (\ref{eq:Hubbard-SB}) that this slave-boson
approximation leads to a picture of quasiparticles with no residual interactions.
The quasiparticle weight reads:
\begin{equation}
Z=z^2 = \frac{1}{n(1-n)}\,\left(\sum_{p=0}^{N-1}\binom{N-1}{p}\phi_p\phi_{p+1}\right)^2
\end{equation}
while the chemical potential shift $\lambda$ in the previous section corresponds in the
present context to the Lagrange multiplier associated with the constraint
(\ref{eq:sb-const1}) (it vanishes at half-filling).

We also note that the non-interacting limit is correctly reproduced, thanks to
the square-root normalization factors \cite{4bosons} introduced in
(\ref{eq:sb-representation}). In that limit, $Z=1$ and $E_0=-N\varepsilon_0$.

In order to analyze the Mott transition, we note that within this approach,  an
insulating solution can be found {\it for arbitrary strength of the coupling $U$}.
It corresponds to the trivial solution:
\begin{equation}
\phi_{N/2}^2\,=\,\binom{N}{N/2}^{-1}\,\,\,,\,\,\,
\phi_p\,=\,0\,\,\,(p\neq \frac{N}{2})
\end{equation}
which yields the $T=0$ occupancies in the atomic limit. This solution is always an
extremum of the variational energy (corresponding to a boundary), but not always a
{\it minimum}. Indeed, a minimum with a lower energy exists for
$\widetilde{U}<\widetilde{U}_{c2}$. To find this critical coupling, we perturb around
the trivial solution. Close to $U_{c2}$, only the occupancies of the
states with charge $N/2\pm 1$ matter, and we set:
$\phi_{N/2}^2=\binom{N}{N/2}^{-1}+\cdots$,
$\phi_{N/2\pm 1}=\delta\phi$, and $\phi_p=\cdots$ otherwise (where $(\cdots)$
denote terms that can be neglected). The energy difference
with the trivial insulating solution reads:
\begin{eqnarray}
\Delta E & = & -4\varepsilon_0\frac{\left(\binom{N-1}{N/2}+\binom{N-1}{N/2-1}\right)^2}
{\binom{N}{N/2}}\,\delta\phi^2\\
\nonumber & + &\frac{\widetilde{U}}{2}
\left(\binom{N}{N/2-1}+\binom{N}{N/2+1}\right)\delta\phi^2
\end{eqnarray}
Hence, the critical coupling at which the metallic solution disappears reads:
\begin{equation}
\widetilde{U}_{c2} = 16\varepsilon_0
\frac{\binom{N-1}{N/2}^2}{\binom{N}{N/2}\binom{N}{N/2-1}}\,=\,
4\varepsilon_0 \,\frac{N+2}{N}
\end{equation}
This value coincides with that obtained by Lu \cite{lu} using the
Gutzwiller approximation. Remarkably, it agrees with our exact
result in the $N\rightarrow \infty$ limit. Together with the
considerations above, this suggests that such a slave boson
approach becomes asymptotically exact, at low-energy and close to
the critical coupling $U_{c2}$, in this limit.
This estimate of the critical coupling is already quite accurate for 
the smallest value of $N=2$ (we take for reference a
semi-circular band with half-width $D=1$):
\begin{equation}
U_c(N=2) = \frac{8D}{3\pi}(N+2) \simeq 3.3
\end{equation}
(the NRG and the projective analysis of the one-orbital model indicate that
$U_c$ is very close to 3). 
On the other hand, the large-$N$ solution given by equation (\ref{uc2-critical}) 
misses obviously the sub-leading term of order $N^0$, and is therefore inaccurate 
at small $N$.

The behaviour of the quasi-particle weight close to the transition can be
obtained by expanding the energy to next order in $\delta\phi$, taking
into account that $\phi_{N/2\pm q}^2 \sim (1-\widetilde{U}/\widetilde{U}_{c2})^q$ (so
that only the terms with $q=0,\pm 1$ matter). This yields:
\begin{equation}\label{QPresidue}
Z \,=\, 1-\frac{U}{U_{c2}} + \cdots
\end{equation}
We note that there are no $1/N$ prefactors in this expression when couplings
are scaled properly.

It is not entirely obvious to decide what the critical coupling
$U_{c1}$ is within this approach. Since the insulating solution
exists for arbitrary $U$, one might be tempted to conclude that
$U_{c1}=0$. Another criterion would be to find at which coupling
the optical gap of the insulator vanishes. Unfortunately, the
above approximation (condensed bosons) is insufficient to discuss
high-energy features (Hubbard bands). Studies that go beyond this
approximation \cite{fluctuations}  and incorporate fluctuations of the slave
bosons suggest that the optical gap closes at $U_{c2}$: in that
sense the slave boson approximation does not yield a coexistence
regime, in contrast to the exact results established above (which are the
generic situation within DMFT).

\section{Finite temperature transitions}
\label{sec:finiteT}

We conclude this paper by addressing the finite-temperature aspects of the transitions
discussed above. As in the one-band case, we expect a coexistence region
$U_{c1}(T)\leq\,U\leq\,U_{c2}(T)$ at finite temperature, which closes at a second-order
(liquid-gas) critical point $(\UC,\TC)$, with $U_{c1}(\TC)=U_{c2}(\TC)=\UC$. Comparison
of free-energies then yield a first-order transition line $T=T_c(U)$ for $T<\TC$.
This has been the subject of many studies in the single
orbital case, and is now established on firm grounds. However, no
analytical estimate of the critical temperature $\TC$ is
available. This is highly desirable, since this scale appears to
be strongly reduced in comparison to the electronic bandwidth. It
would also be interesting to know how this scale depends on the
orbital degeneracy.

Let us start with a qualitative argument, which has
two merits in our view: (i) it allows to understand why $\TC/t$ is such a small
energy scale for a small number of orbitals and (ii) it provides a {\it lower} bound
on $\TC$.
This argument is based on a comparison of the energy difference
between the metallic and insulating solutions at $T=0$, to the
corresponding entropy difference.
For $U=N\widetilde{U}$ close to $U_{c2}$, the energy difference
$\Delta\,E_0\equiv E_I(T=0)-E_M(T=0)$ is expected to behave as:
\begin{equation}
\label{estimenerg}
\Delta E_0 = N\,\frac{(\widetilde{U}_{c2}-\widetilde{U})^2}{\widetilde{U}_{c2}} +\cdots
\end{equation}
This is indeed supported by the slave-boson calculation of Sec.\ref{sec:slave}.
The entropy difference $\Delta S=S_I-S_M$ is dominated by the entropy of the
insulating solution (which has a degenerate ground-state) since
the entropy of the metal vanishes linearly with $T$ at low
temperature $T$. Therefore:
\begin{equation}
\Delta S_0  \sim \ln \binom{N}{N/2} \sim N\,\ln 2
\end{equation}
This allows to estimate the behaviour of the first-order critical line
$U_{c}(T)$ (or $T_c(U)$) at low-temperature (i.e for $U$ close to
$U_{c2}$). Indeed, the transition into the insulating state upon raising
temperature occurs when the entropic gain overcomes the energy cost, leading to:
\begin{equation}\label{Tc_closetouc2}
T_c(U) \simeq \frac{(\widetilde{U}_{c2}-\widetilde{U})^2}{\widetilde{U}_{c2}}
\end{equation}
This estimate is valid only at very low temperature, close to the $T=0$ critical
point $U=U_{c2}$. It indicates a quadratic dependence of the critical line, with a
prefactor which is independent of $N$ \cite{spinodal}.

When $U_{c1}$ is close to $U_{c2}$, which is the case for a small number of orbitals
(in particular in the one-band case), one can roughly estimate the critical endpoint $\TC$
by using (\ref{Tc_closetouc2}) in the regime of couplings where the insulating solution
disappears, leading to:~$\TC\simeq\,(\widetilde{U}_{c2}-U_{c1}/N)^2/\widetilde{U}_{c2}$.
It is clear from this formula that the ratio
$\TC/t$ is small because of the small energy difference between the metallic and insulating
solutions. For a larger number of orbitals however, this can only produce a
{\it lower bound} on the critical endpoint $\TC$. Indeed, the energy difference close to $\TC$
is certainly underestimated by (\ref{estimenerg}), while the entropy difference is reduced as
compared to $N\ln 2$ since the entropy of the metal cannot be neglected at $\TC$ (we note that
it behaves as $\gamma\,T$ with $\gamma\propto N$).
This argument shows that $\TC/t$ must either saturate to a constant, or increase with $N$,
as $N$ increases.

We now turn to a more quantitative study of these finite-temperature issues, using the
Landau functional formalism introduced in \cite{Landau-functionnal}.
A functional of the effective hybridisation is
introduced, which reads, for the $N$-orbital model on the Bethe lattice (semi-circular
d.o.s):
\begin{equation}
\label{eq:defF}
F\left[\Delta\right]\,=\,\frac{N}{2t^2}\,\frac{1}{\beta}\sum_n \Delta(\iomn)^2 \,+\,
\ln Z_{imp} [\Delta]
\end{equation}
in which $Z_{imp}$ is the partition function $Z = \mbox{Tr}\,e^{-S_{imp}}$ of the effective
impurity model, defined by the action:
\begin{eqnarray}
\label{eq:defSimp}\nonumber
S_{imp}& = &\int_0^\beta d\tau\,
\sum_\sigma d^+_\sigma\partial_\tau d_\sigma +
\frac{U}{2} \left[\sum_\sigma(d^+_\sigma d_\sigma-n)\right]^2 +\\
& & + \int_0^\beta d\tau \int_0^\beta d\tau' \Delta(\tau-\tau')
\sum_\sigma d^+_\sigma(\tau) d_\sigma(\tau')
\end{eqnarray}
This functional is locally stationary for these hybridisation functions
which satisfy the DMFT equations:
\begin{equation}
\frac{\delta F}{\delta\Delta} = 0 \,\Leftrightarrow\,
\Delta(\iomn) = t^2 G(\iomn)
\end{equation}
The local stability of these DMFT solutions is controlled by the matrix of
second derivatives:
\begin{equation}\label{eq:stabmatrix}
\chi_{nm} \equiv \frac{\delta^2 F}{\delta\Delta(i\omega_n)\delta\Delta(i\omega_m)}\,
=\, \frac{N}{t^2} + K_{nm}
\end{equation}
with, using (\ref{eq:defF},\ref{eq:defSimp}):
\begin{eqnarray}\nonumber\label{eq:Knm}
K_{nm}\,=\,&\frac{1}{\beta^2}
\int_0^\beta\cdots\int_0^\beta\, d\tau_1 d\tau_1'd\tau_2 d\tau_2'\,
e^{i\omega_n(\tau'_1-\tau_1)+i\omega_m(\tau'_2-\tau_2)}\\
\nonumber
&\sum_{\sigma_1\sigma_2}
\bigl[\langle T_\tau\,d^+_{\sigma_1}(\tau_1)d_{\sigma_1}(\tau'_1)
d^+_{\sigma_2}(\tau_2)d_{\sigma_2}(\tau_2')\rangle - \\
& \langle T_\tau\, d^+_{\sigma_1}(\tau_1)d_{\sigma_1}(\tau'_1)\rangle
 \langle T_\tau\,d^+_{\sigma_2}(\tau_2)d_{\sigma_2}(\tau_2')\rangle
\bigr]
\end{eqnarray}
A solution of the DMFT equations is locally stable provided the $\chi_{nm}$ matrix has
no negative eigenvalues when evaluated for this solution. Hence, the couplings
$U_{c1}(T)$ (resp. $U_{c2}(T)$) correspond to the instability line of the $(U,T)$ plane
along which a negative eigenvalue appears when $\chi_{nm}$ is evaluated for the insulating
(resp. metallic) solution.
For $T=0$, these instability criteria should coincide with those derived from the projective
method described above (as verified below for $U_{c1}(T=0)$).
The critical endpoint $T=\TC$ is such that, for
$T>\TC$, no negative eigenvalue is found, at any coupling $U$, when the stability matrix
is evaluated on the (unique) DMFT solution. It should thus be noted that, in order to
determine $(\UC,\TC)$, it is not necessary to know {\it both} $U_{c1}(T)$ and
$U_{c2}(T)$: either one of them is in principle sufficient.

The practical difficulty in performing this stability analysis at finite temperature
is that: (i) it requires a knowledge of the finite-T solution of the DMFT equations
-or at least a reasonable approximation to it- and (ii) the two-particle correlator
(\ref{eq:Knm}) must be evaluated for this solution. Completing this program, for
arbitrary orbital degeneracy, using numerical methods, is beyond the scope of this paper.
However, we would like to present here a simpler calculation which gives some
insight in the dependence of $U_{c1}(T)$ (and of $\TC$)
on the orbital degeneracy $N$. What we have done is to use the atomic limit as a very rough first
approximation to the insulating solution. We have evaluated the 2-particle correlator
$K_{nm}$ and the stability matrix $\chi_{nm}$ in this limit, for arbitrary $N$. This can
be directly implemented on the computer, using the
spectral decomposition of $K_{nm}$ onto atomic eigenstates (some details are provided
in Appendix \ref{atomic_code}). The stability matrix is then diagonalised numerically
(a truncation to a -large- number of Matsubara frequencies is made), and we search
for the line in the $(U,T)$ plane where a negative eigenvalue is first found.

\begin{figure}
\centerline{\includegraphics[width=9cm]{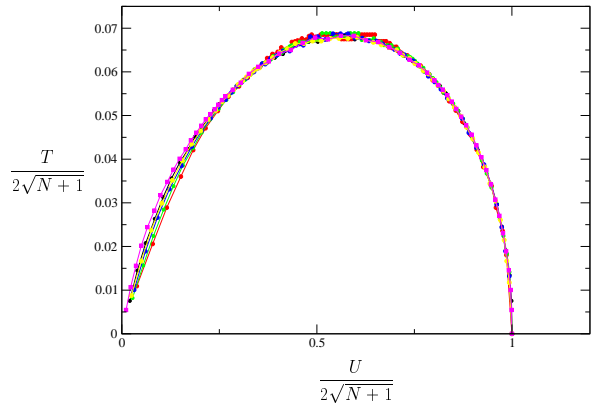}}
\caption{Instability temperature below which a negative eigenvalue is found
when the stability matrix $\chi_{nm}$ of Eq.~(\ref{eq:stabmatrix}) is evaluated in the
atomic approximation. The various curves are for increasing number of orbitals ($N=2,4,6,8,10,20$),
and an almost perfect scaling with $T/\sqrt{N+1}$ and $U/\sqrt{N+1}$ is found.}
\label{fig:Tc_atomic}
\end{figure}

The result of this calculation (for the half-filled case $n=1/2$)
is depicted in Fig.~\ref{fig:Tc_atomic}. There, the
temperature below which a negative eigenvalue is found is plotted as a function of $U$.
We first observe that no instability is found above a critical value of $U$, which does coincide with
the value (\ref{ReducedUCondition}): $U_{c1}(T=0)/t\,\simeq 4\sqrt{N+1}$ estimated above
from the projective method (in the atomic approximation). This is a non-trivial consistency check on
our calculations. Secondly, we find that the instability lines for different values of the orbital
degeneracy $N$ can all be collapsed on a single curve when both $T$ and $U$ are rescaled by
$\sqrt{N+1}$. This can be understood from the fact that, in the atomic limit, only the two scales
$U$ and $T$ enter the two-particle correlator $K_{nm}$ ({\it not} $t$), and that $K$ is of order
$N^2$, so that the instability criterion can be written (from dimensionality considerations):
$\frac{\widetilde{U}}{t} = \Phi(\frac{T/\sqrt{N+1}}{\widetilde{U}})$.
Naturally, we do not expect the curve in Fig.~\ref{fig:Tc_atomic} to be a quantitatively reliable
determination of $U_{c1}(T)$, because of the atomic approximation involved. In particular, we see that,
instead of terminating at a critical endpoint $(\UC,\TC)$, the instability line in Fig.~\ref{fig:Tc_atomic}
bends back, yielding a stability window at small $U$ which is certainly a spurious aspect of the
approximation. Hence only the branch of the curve corresponding to larger values of $U$ has physical
significance.

It is tempting to conclude, from the the observed scaling with $\sqrt{N+1}$, that $\TC$
itself will grow in that manner as $N$ is increased.
Indeed, we note that recent QMC calculations by Amadon and Biermann~\cite{Amadon-QMC}
do show a marked increase of $\TC$ with $N$, not inconsistent with a $\sqrt{N}$ scaling.
However, we do not consider this issue to be entirely settled: rather, this atomic estimate
provides an {\it upper bound} on the growth of $\TC$ with $N$. Indeed, it neglects the
presence of the quasiparticle resonance (which is reduced but still present as
temperature is raised close to $\TC$). The resonance makes the solution less stable, so that
the atomic calculation is likely to overestimate $\TC$. Combined with the above entropy argument
(which provides a lower bound), we conclude that $\TC/t$ is an increasing function of $N$ at
moderate values of $N$, which either saturates at large-$N$ or increases at most as $\sqrt{N}$.
Clearly, a more refined analysis of the above stability condition is required to settle this
issue.

\section{Conclusion}

In this paper, we have studied the Mott transition of the
$N$-orbital Hubbard model in the fully symmetric case. The physical
picture is qualitatively the same as for $N=2$, {\it i.e.}
a coexistence region exists between an insulating and a metallic
solution within two distincts critical couplings $U_{c1}$ and
$U_{c2}$. We have shown that these critical couplings {\it do not have  the
same dependence on the number of orbitals}, as the latter becomes large:
$U_{c1}\propto\sqrt{N}$ while $U_{c2}\propto N$. We have obtained
an exact analytical determination of the critical value
$\widetilde{U}_{c2}=U_{c2}/N$ in this limit. These results explain
the widening of the coexistence window $[U_{c1},U_{c2}]$ observed
in the numerical simulations of Ref.~\cite{multiband-QMC-marcelo},
and more recently in \cite{Amadon-QMC}. Our findings also put
the results of Refs.~\cite{multiband-QMC-gun,multiband-QMC-han} in a new
perspective, as discussed in the introduction.

We have shown that, at large $N$ and close to the critical coupling $U_{c2}$,
the scaling $U=N\widetilde{U}$ is appropriate, so that the separation of scale
which is at the heart of the projective method becomes asymptotically exact.
The low-energy physics close to $U_{c2}$ is then well described by slave-boson
like approximations. This is the case in particular for the quasi-particle
peak in the spectral function, which is separated from Hubbard bands by a large
energy (of order $N\widetilde{U}$) and is accurately described by the simple
form (\ref{eq:Ldosfinal}).
In view of practical applications to electronic structure calculations
with large orbital degeneracy (e.g f-electron systems), it would be highly
desirable to have also an analytical determination of the high-energy part of the
spectral function in the large-$N$ limit (or at least reasonable approximations to it).
This is left for future investigations.

Another issue that we have only partly addressed is the finite-temperature aspects
of these transitions. The whole finite-temperature coexistence
region widens as $N$ is increased, and the critical temperature
associated with the Mott critical endpoint does increase with the number of orbitals
at intermediate values of $N$.
The precise behaviour of $\TC/t$ at large $N$ deserves further studies however.

\begin{acknowledgments}
We thank B. Amadon and S. Biermann for useful 
discussions, and for sharing with us their recent QMC numerical results
for this model.
This work was  supported by the NSF under grant
NSF DMR 0096462 and by  the PICS program of the CNRS (under contract PICS 1062).   
\end{acknowledgments}

\appendix
\section{Large-$N$ scalings and the atomic limit}
\label{appendix}

In this appendix, we provide a detailed proof that the corrections in $\VH$ to the
atomic limit ($\VH=0$) do not modify the large-$N$ dependence of the critical couplings.
This is done by an explicit investigation of the structure of these corrections.
Moreover, we show that for $U\propto N$ (i.e when $U_{c2}$ is considered), these corrections
only produce subdominant terms which can be ignored at large-$N$.

\subsection{Corrections in $\VH$ for $U=N\widetilde{U}$.}

For our purpose, the operator $H_{\mr{H}}$ can be written :
\begin{eqnarray}\label{rewriteHH}
H_{\mr{H}} & = & H_{\mr{H}}^0 + H_{\mr{H}}^1 \\
\nonumber
H_{\mr{H}}^0 & = & \sum_{\kH, \s} \ek \ckc \ck  +
\frac{U}{2} \left[ \sum_{\s} (\dsc \ds - n) \right] ^2 \\
\nonumber
H_{\mr{H}}^1 & = &  \VH\sum_{\sigma } \sum_{k} \ckc \ds + \dsc \ck
\end{eqnarray}

First let us consider the ground state energy $\mr{E_{gs}}$.
 As the lower Hubbard band $\mr{H}^-$ is filled in
$\ket{\mu}_0 $, we prefer to remove the constant contribution
$\sum_{k\in\mr{H}^- , \s} \ek$, so that the zeroth order energy is
$\mr{E}_0 = \,_0\!\bra{\mu} H_{\mr{H}}^0 \ket{\mu}_0 = 0$.
The next order term $\delta \mr{E}_1 = \,_0\!\bra{\mu} H_{\mr{H}}^1
\ket{\mu}_0 $ is zero by conservation of the total charge $Q$ on the
$d$-level, so we need to look at the second order contribution:
\begin{eqnarray}
\delta \mr{E}_2 & = & \sum_{\ket{\psi}} \frac{ \left|
\bra{\psi} H_{\mr{H}}^1 \ket{\mu}_0 \right|^2}
{\mr{E}_0 - \mr{E}_{\psi}}\\
\nonumber
& = & - (N-Q) \sum_{k\in\mr{H}^-} \frac{\VH^2}{\frac{U}{2}-\ek}
- Q \sum_{k\in\mr{H}^+} \frac{\VH^2}{\frac{U}{2}+\ek}
\end{eqnarray}
since $\ket{\psi}$ can only possess a charge $Q+1$ (resp. $Q-1$) for the $d$-electron
and a hole (resp. particule) excitation in the bath.
Clearly as $\ek$ is negative (positive) for $k \in \mr{H}^-$ ($\mr{H}^+$), the
denominators are of order $N$ (since $U=N \widetilde{U}$) and
$\delta \mr{E}_2$ is therefore scaling as $N^0$ at large $N$.
We would like to infer this to be true at all order in the developement
in $\VH$. One can easily classify into two categories the intermediate states
that appear at order $\VH^4$ and beyond. There can either be a charge
different form $Q$ on the impurity (as in the previous computation at order
$\VH^2$) and then the denominator coming from this state provides
(at least ) a factor $N^{-1}$ by the presence of the Coulomb
energy $U=N\widetilde{U}$.
Or the impurity can be found in a charge $Q$ state (with possibly
a reorganization of the spin configuration) without a Coulomb
energy cost. However this state possesses also many particle-hole excitations
in the bath, and this leads to a denominator which scales as the Mott gap
$\Delta_g$. For example let us consider such a ket generated at order
$\VH^4$, like
$\ket{\s_1 \ldots \s_Q}_d \ket{2 \ldots N}_{-} \ket{1}_{+}$.
This particular state (but the argument is general), contributes a factor
$(\ek - \epsilon_{k'})^{-1}$
where $k \in \mr{H}^-$ and $k' \in \mr{H}^+$. This quantity is always
smaller than $\Delta_g^{-1}$ because the Mott gap $\Delta_g$ is also the
gap of the $\ck$ electrons (through the self-consistency).
As the gap $\Delta_g$ is expected to be of order $U=N\widetilde{U}$
when $U$ is close to $U_{c2}$, each particle-hole excitation also
provides a factor $N^{-1}$ in the perturbative expansion.
This is enough to conclude that, at a given order in perturbation theory,
each of the denominator gives a factor $N^{-1}$ that balance the
combinatorial factor coming from the string of charge excitation on the
impurity level. Thus $\mr{E}_{gs}$ is at most of order 1 when $N$ is large.

We finally need to compute $J^{\s \s' }_{ \mu  \nu }$ at large $N$.
Let us consider first the case $\mu \neq \nu$, so that $\mu$ and
$\nu$ will differ by only one spin flip $\s \leftrightarrow \s'$. We
can now calculate at the lowest order in $\VH$ the desired matrix element:
\begin{equation}
\Gamma \equiv \langle\mu| d_{\s }
\frac{1}{H_{\mr{H}}^0 + H_\mr{H}^1 -\mr{E}_{gs}} d^{\dagger}_{\s'} |\nu\rangle
\end{equation}
with
\begin{eqnarray}
\ket{\mu} & \equiv & \ket{\mu}_0 + \ket{\mu}_1 \\
\nonumber
\ket{\mu}_1 & = &
\!\!\!\!  \sum_{k\in\mr{H}^-,\s} \frac{\VH}{\frac{U}{2}-\ek} \, \dsc\ck \ket{\mu}_0
+\!\!\!\! \sum_{k\in\mr{H}^+,\s} \frac{\VH}{\frac{U}{2}+\ek} \, \ckc\ds \ket{\mu}_0
\end{eqnarray}
The contribution at order $(\VH)^0$ is therefore:
\begin{equation}
\delta \Gamma_0 = \,_0 \langle\mu| d_{\s }
\frac{1}{H_{\mr{H}}^0 -\mr{E}_{gs}} d^{\dagger}_{\s'}
| \nu \rangle_0
%= \frac{1}{\frac{U}{2} - \mr{E}_{gs}}
\sim \frac{2}{U}
\label{eq:Gamma}
\end{equation}
because $\mr{E}_{gs} = {\mathcal{O}}(1)$ and $U = N \widetilde{U}$.
All terms coming at order $\VH$ cancel by conservation of the charge, so we
examine now the next leading order for $\Gamma$, $\delta\Gamma_2$, which is
composed of three terms: one from the correction by $\VH$ to each of the
two external kets,
one from the development of the denominator in (\ref{eq:Gamma}) at
order $\VH^2$, and a mixed term (between only one ket correction and the
denominator developed at first order).
Let us examine the first of these second order contributions:
\begin{equation}
\delta \Gamma_2^{(1)} = \,_1 \langle\mu| d_{\s }
\frac{1}{H_{\mr{H}}^0 -\mr{E}_{gs}} d^{\dagger}_{\s'}
| \nu \rangle_1
\end{equation}
The state $d^{\dagger}_{\s'}| \nu \rangle_1 $ is composed of
charge $Q+2$ excitations (with energy $2^2 U/2$) and of
particle-hole excitations in the bath with a charge $Q$ on the impurity
(this has the energy $\ek - \epsilon_{k'}$). Both energies scale like $N$
and contribute an overall factor $N^{-3}$ to $\delta \Gamma_2^{(1)}$.
There is also a combinatorial factor of order $N$ coming from the
choice in the spin-excitation in $|\nu \rangle_1 $, but there is none from the
state $|\mu \rangle_1 $ as the spin configuration in
$d^{\dagger}_{\s'}| \mu \rangle_1 $
is fixed by the one in $d^{\dagger}_{\s'}| \nu \rangle_1 $.
Therefore $\delta \Gamma_2^{(1)} = {\mathcal{O}}(N^{-2})$. The same argument works
for the two other contribution at order $\VH^2$.

More generally, at a given order $p$ in the development in $\VH$,
each of the $p+1$ denominators provides a factor of order $N^{-1}$, because
there is a finite gap $\Delta_g \sim U_{c2} \propto N$ between
the two Hubbard bands \cite{AG-GK-Comment}. However the string of
$p$ spin-flips generated by the $H_\mr{mix}$ terms only
contributes to an overall combinatorial factor at most $N^{p/2}$ because one has
to connect two fixed external configurations $\ket{\nu}$ and $\ket{\mu}$.
Hence in the limit of large $N$ all terms after the first one are
sub-dominant, so that:
\begin{equation}
\Gamma = \delta \Gamma_0 + {\mathcal O} \left(
\frac{1}{N^2} \right) = \frac{2}{U} = \frac{2}{N \widetilde{U}}
\end{equation}
meaning that the atomic result for $\Gamma$ is exact in the large-$N$ limit {\sl when
$U\propto N$}.

At this stage, we want to stress {\it a posteriori} that the
hypotheses made in the preceding argument are consistent with this
result. First it is correct to fix the chemical potential $\mut=0$
from the atomic limit ($\VH=0$), because perturbative corrections do not
change the occupancy of the $d$-orbital, as can be checked from a direct
expansion in $\VH$ of $\left<\sum_\sigma d^\dagger_\sigma \ds \right>$.
Finally the existence of a large Mott gap $\Delta_g \propto N$ can also be justified on
the same grounds.

\subsection{Some remarks about $U_{c1}$}\label{sub:Uc1}

When $U\propto \sqrt{N}$ on the other hand, the perturbative series for
$\Gamma$ does not appear to stop after the first contribution, showing that
$U_{c1}$ is affected by $\VH$ in the large $N$ limit.
It is however straightforward to follow the line of arguments given in
Section \ref{sub:limitatomicexact} to see that each term in this
development provides a leading contribution of order $1/\sqrt{N}$
( using $U \sim \Delta_g \propto \sqrt{N}$). $\Gamma$ can not be computed
exactly in that case, because
$B(N,n;\VH/(\sqrt{N}\widetilde{U})) $ is now a non trivial function
of $\VH/\widetilde{U}$ at $N=\infty$.
But this is enough to conclude that $B(N,n;\VH/U)$ is of order $1$ at large $N$ when
$U\propto \sqrt{N}$, so that the result $U_{c1} \sim \sqrt{N} \widetilde{U}_{c1}$
holds. However, an exact evaluation of $\widetilde{U}_{c1}$ appears to be
a difficult task, even at $N=\infty$.

\section{Numerical calculation of the stability matrix in the atomic limit}
\label{atomic_code}

In this Appendix, we provide details on the computation of the stability matrix
$\chi_{nm}$ in the $N$-orbital atomic limit (cf. Fig.~\ref{fig:Tc_atomic}).
First, we use a spectral decomposition over the atomic eigenstates in Eq.~(\ref{eq:Knm}).
This yields:
\begin{equation*}
K_{nm}= A (i\omega_{n},-i\omega_{n},i\nu_{n},-i\nu_{n})/Z_{at} -
B (i\omega_{n}) B (i\nu_{n})/Z_{at}^{2}
\end{equation*}
with
\begin{equation*}
Z_{at}= \sum_{Q=0}^{N} e^{-\beta E_{Q}} \binom{N}{Q}
\end{equation*}
\begin{widetext}
\begin{align*}
&A (\omega_{j},j=1..4 )\equiv
N\sum_{
\atop {0\leq Q\leq N} {P\in {\cal  S}_{4}}
}
 e^{-\beta E_{Q}} \epsilon (P)
\Bigl ( C_{1} (Q,P)  +  (N-1)C_{2} (Q,P)
\Bigr )
F_{4} ( \omega_{P (j)} + \Delta_{Q+ \sum_{i=j+1}^{4} a_{P(i)}, a_{P (j)}}
, j=1..4)
\\
&B (i\omega_{n}) \equiv N \sum_{Q=0}^{N} e^{-\beta E_{Q}}
\left[
\binom{N-1}{Q-1} F_{2} (i \omega_{n} + \Delta_{Q-1,1},-i\omega_{n}+ \Delta_{Q,-1})
-
\binom{N-1}{Q} F_{2} (-i \omega_{n} + \Delta_{Q+1,-1},i\omega_{n}+ \Delta_{Q,1})
\right]
\\
&C_{2} (Q,P) \equiv
\begin{cases}
\epsilon (P) \binom{N-2}{Q-2} \quad \text{if} \quad P^{-1} (1) <P^{-1} (2) \text{\quad and\quad }
 P^{-1} (3) <P^{-1} (4) \\
\epsilon (P) \binom{N-2}{Q} \quad \text{if} \quad P^{-1} (1) >P^{-1} (2) \text{\quad and\quad }
P^{-1} (3) >P^{-1} (4) \\
-\epsilon (P) \binom{N-2}{Q-1 } \quad \text{else}
\end{cases}
\end{align*}
\end{widetext}
\begin{align*}
&C_{1} (Q,P) \equiv
\begin{cases}
\binom{N-1}{Q-1} \quad \text{if} \quad P^{-1} (\{1,3 \}) = \{1,3 \} \\
\binom{N-1}{Q} \quad \text{if} \quad P^{-1} (\{1,3 \}) = \{2,4 \} \\
0 \quad \text{else}
\end{cases}
\\
&F_{n} (\omega_{j},j=1..n) \equiv
\int_{\beta >\tau_{1}> \dots >\tau_{n}}
\left(\prod_{i=1}^{n} d\tau_{i} \right)
\exp \Bigl ( \sum_{i=1}^{n} \omega_{j} \tau_{j}     \Bigr )
\end{align*}
In these expressions, $E_{Q}\equiv U(Q-N/2)^{2}/2$ are the atomic energy levels,
$\Delta_{Q,a} \equiv E_{Q+a} - E_{Q}$, $\epsilon $ is the signature of the permutation $P$,
and we used the notation $f (x_{j},j=1..4)\equiv f (x_{1},x_{2},x_{3},x_{4})$.
We compute $F_{4}$ and $F_{2}$,
using the relations $\omega_{1} + \omega_{2} + \omega_{3} + \omega_{4}=0$ and
$\omega_{1} + \omega_{2}=0$ respectively.
The algorithm can be decomposed into three functions :
{\sl (i)} From $U$ and $T$, compute $\chi_{nm}$ for $|n|,|m|<100$ and then its lowest eigenvalue
$E_{0} (U,T)$;
{\sl (ii)} Find a point $p0$ below the instability temperature $T_{i}$
 and a majoration  $T_{M}$ of $max (T_{i})$ and $U_{M}$ of $U_{c1}$; define a path in the $(U,T)$
from $(0,0)$ to $(0,T_{M})$ to $(U_{M},T_{M})$ to $(U_{M},0)$;
{\sl (iii)} For about $50$ points $p$ on this path, solve for $E_{0} (U,T)=0$ on the line defined by
$p$ and $p_{0}$ using a dichotomy.
To speed up the computation, the sums over permutations and $Q$ in $A$ and $B$ are
expanded automatically into C++ code, which is inlined in the main program. Diagonalisations are performed
using the Fortran LAPACK library. Codes will be available at www-spht.cea.fr/$\tilde{\ }$parcolle/.


\begin{thebibliography}{99}

\bibitem{imada-review} For a review, see e.g:
M. Imada, A. Fujimori and Y. Tokura, {\sl Rev. Mod. Phys.}
{\bf 70}, 1039 (1998).
%
\bibitem{gunnarsson-fullerenes} For a review, see e.g:
O. Gunnarsson, {\sl Rev. Mod. Phys.} {\bf 69}, 575 (1996).
%
\bibitem{lefebvre} S. Lefebvre, P. Wzietek, S. Brown, C. Bourbonnais, D. J{\'e}rome,
C. M{\'e}zi{\`e}re, M. Fourmigu{\'e}, and P. Batail {\sl Phys. Rev. Lett.} {\bf 85},
5420-5423 (2000).
%
\bibitem{held-review} For a review, see e.g:
%K. Held, I.A. Nekrasov, G. Keller, 
%V. Eyert, N. Bl{\"u}mer, A.K. McMahan, R.T. Scalettar, T. Pruschke,
%V.I. Anisimov and D. Vollhardt, {\sl cond-mat} 0112079
%
Strong Coulomb Correlations in Electronic Structure Calculations,
Edited by V. Anisimov. Gordon and Breach (2001)

\bibitem{large-d} W. Metzner and D. Vollhardt, Phys. Rev. Lett.  {\bf  62} 324 (1989).
%
\bibitem{IPT} A. Georges and Kotliar, Phys. Rev. B {\bf 45}, 6479 (1992).
%
\bibitem{RMP} For a review, see: A. Georges, G. Kotliar, W. Krauth and M. Rozenberg,
Rev. Mod. Phys. {\bf 68}, 13 (1996).
%
\bibitem{pruschke-review} For a review, see: T. Pruschke, M. Jarrell, and J.K. Freericks, {\sl Adv. Phys.}
{\bf 42}, 187 (1995)
%
\bibitem{ref-QMC}  M. Jarrell, Phys. Rev. Lett. {\bf 69}, 168 (1992);
M. Rozenberg, X. Y. Zhang and G. Kotliar, Phys. Rev. Lett. {\bf 69},
1236 (1992); A. Georges and W. Krauth, Phys. Rev. Lett, {\bf 69}, 1240 (1992).
%
\bibitem{Mott_scenario} X.Y. Zhang, M. J. Rozenberg, and G. Kotliar
{\sl Phys. Rev. Lett.} {\bf 70}, 1666, (1993); A. Georges and W. Krauth
{\sl Phys. Rev. B} {\bf 48}, 7167 (1993); M. Rozenberg, G. Kotliar, and X. Y. Zhang
{\sl Phys. Rev. B} {\bf  49}, 10181 (1994).
%
\bibitem{optics}
M. Rozenberg G. Kotliar H. Kajuter G. A. Thomas D. H. Rapkine J. M.
Honig and P. Metcalf, Phys.  Rev.  Lett. {\bf 75}, 105  (1995). 

\bibitem{Moeller} G. Moeller, Q. Si, G. Kotliar, M. Rozenberg
and D.S. Fisher, Phys. Rev. Lett. {\bf 74}, 2082 (1995).
%
\bibitem{MoellerPHD} G. Moeller, { \it Ph.D Rutgers}, unpublished.\\
http://www.physics\-.rutgers.edu/$\tilde{\ }$kotliar/thesis/moel\-ler.ps.gz.
%
\bibitem{Landau-functionnal} G. Kotliar, Eur. Phys. J. B, {\bf 11}, 27 (1999).
G. Kotliar, E. Lange and M.J. Rozenberg, Phys. Rev. Lett.
{\bf 84}, 5180 (2000). 
%
\bibitem{QMC-oudo} J. Joo and V. Oudovenko {\sl Phys. Rev. B} {\bf 64}, 193102 (2001).
%
\bibitem{Bulla-NRG} R. Bulla, Adv. Solid State Phys. {\bf 40},
169 (2000); R. Bulla, T. Costi and D. Vollhardt {\sl Phys. Rev. B} {\bf 64}, 045103 (2001).
%
\bibitem{exact-diag} M. Caffarel and W. Krauth, Phys. Rev. Lett. {\bf 72},
1545 (1994);
Q. Si, M. Rozenberg, G. Kotliar and A. Ruckenstein, Phys. Rev. Lett. {\bf 72}, 2761 (1994).
%
\bibitem{dos} This applies provided the bare density of states has a sufficiently
smooth variation with energy, which is the case of the model d.o.s considered in this paper
%
\bibitem{multiband-QMC-gun}
O. Gunnarsson, E. Koch and R. Martin, Phys. Rev. B {\bf 54},
R11026 (1996).
%
%
\bibitem{multiband-QMC-han}
J. E. Han, M. Jarrell and D. L. Cox, Phys. Rev. B {\bf 58}, R4199
(1998).
%
\bibitem{multiband-QMC-marcelo}
M. J. Rozenberg, Phys. Rev. B {\bf 55}, R4855 (1997).
%
\bibitem{Amadon-QMC} B. Amadon and S. Biermann, unpublished.
%
\bibitem{Kajueter-degeneracy} 
G. Kotliar and H. Kajueter, Phys. Rev. B 54, R14221 (1996).

H. Kajueter and G. Kotliar,
Int. J. Mod. Phys. {\bf 11}, 729 (1997).
%
\bibitem{lu} P. Lu, Phys. Rev. B {\bf 49}, 5687 (1994).
%
\bibitem{AG-GK-Comment}A. Georges and G. Kotliar, Phys. Rev. Lett.
{\bf 84}, 3500 (2000).
%
\bibitem{Read-Newns} N. Read and D.M. Newns, Adv. Phys. 36 p.799 (1987).
%
\bibitem{4bosons} G. Kotliar and A. Ruckenstein, Phys. Rev. Lett {\bf
52}, 1362 (1986).
%
\bibitem{fresard-kotliar} R. Fresard and G. Kotliar, Phys. Rev. B {\bf
56}, 12909 (1997).

H. Hasegawa, J. Phys. Soc. Jpn. {\bf 66}, 3522 (1997).

%
\bibitem{fluctuations} R. Raimondi and C. Castellani, Phys. Rev. B {\bf48},
11453 (1993).
%
\bibitem{spinodal} One should note however that the large-$N$ solution of Sec.~\ref{Sec.NGrand}
suggests that the temperature above which the
metallic solution disappears (spinodal) has a different temperature dependence:
$T\propto (\widetilde{U}-\widetilde{U}_{c2})^{3/2}$.

\end{thebibliography}
\end{document}